\documentclass{article}
\usepackage{dcase2018,amsmath,graphicx,url,times,booktabs, tabularx}
\usepackage[utf8]{inputenc}
\usepackage{graphicx}
\usepackage{caption}
\usepackage{subcaption}

\usepackage{nameref,hyperref}

\title{Data-Efficient Weakly Supervised Learning for Low-Resource Audio Event Detection Using Deep Learning}

\twoauthors
  {Veronica Morfi}
    {Machine Listening Lab,\\
	Centre for Digital Music (C4DM),\\
	Queen Mary University of London, UK\\
     g.v.morfi@qmul.ac.uk}
  {Dan Stowell}
    { Machine Listening Lab,\\
	Centre for Digital Music (C4DM),\\
	Queen Mary University of London, UK\\
     dan.stowell@qmul.ac.uk}

\begin{document}

\ninept
\maketitle

\begin{sloppy}

\begin{abstract}
We propose a method to perform audio event detection under the common constraint that only limited training data are available. In training a deep learning system to perform audio event detection, two practical problems arise. Firstly, most datasets are ``weakly labelled'' having only a list of events present in each recording without any temporal information for training. Secondly, deep neural networks need a very large amount of labelled training data to achieve good quality performance, yet in practice it is difficult to collect enough samples for most classes of interest. In this paper, we propose a data-efficient training of a stacked convolutional and recurrent neural network. This neural network is trained in a multi instance learning setting for which we introduce a new loss function that leads to improved training compared to the usual approaches for weakly supervised learning. We successfully test our approach on two low-resource datasets that lack temporal labels.
\end{abstract}


\section{Introduction}
\label{sec:intro}

In recent decades, there has been an increasing amount of audio datasets that have labels assigned to them to indicate the presence or not of a specific event type. This is related to tagging of audio recordings \cite{Kong:17, Xu:17b, Xu:17a, Adavanne:17a}. However, in many cases these labels do not contain any information about the temporal location of each event or the number of occurrences in a recording. This type of label, which we will refer to as \textit{weak}, lacks any temporal information. Collecting and annotating data with strong labels, labels that contain temporal information about the events, is a time consuming task involving a lot of manual labour. On the other hand collecting weakly labelled data takes much less time, since the annotator only has to mark the active sound event classes and not their exact boundaries. 

In comparison to supervised techniques trained on strong labels, there has been relatively little work on audio event detection using weakly labelled data. In \cite{Briggs:12, Ruiz:15, Fanioudakis:17} the authors try to exploit weak labels in birdsong detection and bird species classification, while in \cite{Schluter:16} singing voice is pinpointed from weakly labelled examples. Furthermore, in \cite{Adavanne:17b} the authors train a network that can do automatic scene transcription from weak labels and in \cite{Hershey:17} audio from YouTube videos is used in order to train and compare different previously proposed convolutional neural network architectures for audio event detection and classification. Finally, in \cite{Kumar:16, Kumar:17} the authors use weakly labelled data for audio event detection in order to move from the weak labels space to strong labels. Most of these methods formulate the provided weak labels of the recordings into a multi instance learning (MIL) problem. 

Machine learning has experienced a strong growth in recent years, due to increased dataset sizes and computational power, and to advances in deep learning methods that can learn to make predictions in extremely nonlinear problem settings \cite{LeCun:15}. However, a large amount of data is needed in order to train a neural network that can achieve a good quality performance. Depending on the audio event to be detected and classified in each task it may become difficult to collect enough samples for them. Annotating data with \textit{strong} labels, labels that contain temporal information about the events, to train audio event detectors is a time consuming process involving a lot of manual labour. On the other hand, collecting weakly labelled data takes much less time, since the annotator only has to mark the active sound event classes and not their exact boundaries. We refer to datasets that only have weak labels, may contain rare events and have limited amounts of training data as \textit{low-resource} datasets.

In this paper, we propose a network that uses low-resource datasets in an efficient way in order to predict audio event detection using only the weak labels provided for each recording during training. This network is trained in a MIL setting where we propose a new loss function that outperforms the most commonly used losses when trying to derive the strong labels from weakly labelled data. The rest of the paper is structured as follows: Section \ref{sec:Background} describes the multi instance learning setting, Section \ref{sec:bigidea} presents our method. Evaluations on two different low-resource datasets follow in Section \ref{sec:Eval}, with the conclusions in Section \ref{sec:discussion}.

\section{Multi Instance Learning}
\label{sec:Background}
Training audio event detectors on low-resource datasets presents the issue of weak-to-strong prediction. Low-resource datasets only provide the user with weak labels that don't include any temporal information about the events but only denote the presence or absence of a specific class in a recording. However, audio event detectors produce labels with start and end times, referred to as strong labels, hence provide full temporal information about the events in a recording. 

The most common way to train a network for weak-to-strong prediction is the multi instance learning (MIL) setting. The concept of MIL was first properly developed in \cite{Dietterich:97} for drug activity detection. MIL is described in terms of \textit{bags}, with a bag being a collection of instances. The existing weak labels are attached to the bags, rather than the individual instances within them. Positive bags have at least one positive instance, an instance for which the target class is active. On the other hand, negative bags contain negative instances only. A negative bag is thus pure while a positive bag is presumably impure, since the latter most likely contains both positive and negative instances. There is no direct knowledge of whether an instance in a positive bag is positive or negative. Thus, it is the bag-label pairs and not the instance-label pairs which form the training data, and from which a classifier which classifies individual instances must be learned.

Let the training data be composed of $N$ bags, i.e. $\left\lbrace B_1, B_2, ..., B_N \right\rbrace$, the $i$-th bag is composed of $M_i$ instances, i.e. $\left\lbrace B_{i1}, B_{i2}, ..., B_{iM_i} \right\rbrace$, where each instance is a $p$-dimensional feature vector, e.g. the $j$-th instance of the $i$-th bag is $\left[ B_{ij1}, B_{ij2}, ..., B_{ijp} \right] ^T$. We represent the bag-label pairs as $\left( B_i, Y_i \right)$, where $Y_i \in \left\lbrace 0, 1 \right\rbrace$ is the bag label for bag $B_i$. $Y_i = 0$ denotes a negative bag and $Y_i = 1$ denotes a positive bag. 

One na\"ive but commonly used way of inferring the individual instances' labels from the bag labels is assigning the bag label to each instance of that bag: we refer to this method as \textit{false strong labelling}. During training, a neural network in the MIL setting with false strong labels tries to minimise the average divergence between the network output for each instance and the false strong labels assigned to them, identically to an ordinary supervised learning scenario. However, it is evident that the false strong labelling approach is an approximation of the loss for a strong label prediction task, hence it has some disadvantages. When using false strong labels some kind of early stopping is necessary since when minimal loss is achieved that would mean all positive instance predictions for a positive bag. However, there is no clear way of defining a specific point for early stopping. This is an issue that all methods in the MIL setting face. 

As an alternative to false strong labels, one can attempt to infer labels of individual instances in bag $B_i$ by making a few educated assumptions. The most common ones are: if $Y_i = 0 $, all instances of bag $B_i$ are negative instances, hence $y_{ij} = 0, \forall j$, while on the other hand, if $Y_i = 1$, at least one instance of bag $B_i$ is equal to one. For all instances of bag $B_i$, this relation between the bag label and instance labels can be simply written as $Y_i = \max_j y_{ij}$. Using this assumption in the MIL setting, we must modify the manner in which the divergence to be minimized is computed, to utilize only weak labels, as proposed in \cite{Zhou:02}. 

Let $o_{ij}$ represent the output of the network for input $B_{ij}$, the $j$-th instance in $B_i$, the $i$-th bag of training instances. We define overall divergence on the training set as the sum of the bag-level divergences $E_i$, each computed for bag $B_i$:

\begin{equation}
E = \sum_{i=1}^N E_i = \sum_{i=1}^N \frac{1}{2} \left( \max_{1 \leq j \leq M_j} (o_{ij}) - Y_i \right) ^2
\label{eq:E_i}
\end{equation}
where $Y_i$ is the label assigned to bag $B_i$. 

This indicates that if at least one instance of a positive bag is perfectly predicted as positive, or all the instances of a negative bag are perfectly predicted as negative, then the error on the concerned bag is zero. Otherwise, the weights will be updated according to the error on the instance whose corresponding actual output is the maximal among all the instances in the bag. Note that such an instance is typically the most easy to be predicted as positive for a positive bag, while it is the most difficult to be predicted as negative for a negative bag. On an instance-level, when using max to compute the loss, only one instance per bag contributes to the gradient, which may lead to inefficient training. In positive bags, the network only has to accurately predict the label for the easiest positive instance to reach a perfect accuracy, thus not paying as much attention to the rest positive instances that might be harder to accurately detect.


In this work, we perform audio event detection by training a neural network using weakly labelled data in a MIL setting. Since the loss function can have a dramatic effect on the utility of MIL training, we propose a loss function at a bag-level that does not have the disadvantages of the above methods and leads to improved training.
\vspace{-0.4cm}
\section{Method}
\label{sec:bigidea}

\subsection{Loss Function}
\label{subsub:loss}
Using all instances in a bag for computing error and backpropagated gradient is important, since the network ideally should acquire knowledge from every instance in each epoch. However, it is hard to find an elegant theoretical interpretation of the characteristics of the instances in a bag. On the other hand, simple assumptions about these characteristics can achieve a similar effect. One approach is to consider the mean of the instance predictions of a bag. If a bag is negative the mean should be zero, while if it is positive it should be greater than zero. The true mean is unknown in weakly labelled data. A na\"ive assumption is to presume that approximately half of the time an event will be present in a recording. Even though this is not true all of the time, it takes into consideration the predictions for all instances, and also inserts a bias to the loss that will keep producing some gradient even after the max term has reached its perfect accuracy. Another simple yet accurate assumption is that on both negative and positive recordings the minimum predictions at an instance-level should be zero. It is possible for a positive recording to have no negative frames however it is extremely rare in practice. This assumption could be used in synergy with max and mean to enforce the prediction of negative instances even on positive recordings and manage a certain level of the bias that is introduced with considering mean in the computation of the loss. 

Our proposed loss function that takes into account all the above mentioned assumptions is computed as:
\begin{multline} 
Loss = \frac{1}{3} \big( bin\_cr(max_j (o_{ij}), Y_i) \\
+ bin\_cr(mean_j (o_{ij}), \frac{Y_i}{2}) + bin\_cr(min_j (o_{ij}), 0) \big)
\label{eq:mmmloss}
\end{multline} 
where $bin\_cr(x,y)$ is a function that computes the binary cross-entropy between $x$ and $y$, $o_{ij}$ are all the predicted strong labels of bag $B_i$, where $j = 1...M_i$ with $M_i$ being the total number of instances in a bag, and $Y_i$ is the label of the bag. 

We refer to this as an \textit{MIL setting using MMM}. For negative recordings, (\ref{eq:mmmloss}) will compute the binary cross-entropy between the max, mean and min of the predictions of the instances of a bag $B_i$ and zero. This denotes that the predictions for all instances of a negative recording should be zero. On the other hand, for positive recordings the predictions should span the full dynamic range from zero to one, biased towards a similar amount of positive and negative instances. Our proposed loss function is designed to balance the positive and negative predictions in a bag resulting in a network that has the flexibility of learning from harder-to-predict positive instances even after many epochs. This is due to the fact that there are no obvious local minima to get stuck in as in the max case.

\subsection{Training Settings}
\label{subsec:training}
As input to our proposed method, log mel-band energy is extracted from audio in 23ms Hamming windows with 50\% overlap. In order to do so the \texttt{librosa} Python library is used.\footnote{\url{https://librosa.github.io/librosa/index.html}} In total, 40 mel-bands are used in the 0--44100 Hz range. For a given 5 second audio input, the feature extraction produces a 432x40 output ($T = 432$).

We use a stacked convolutional and recurrent neural network architecture (cf. \cite{Adavanne:17b}) to predict the strong labels of a recording. Table \ref{tab:when} presents the overall architecture of the proposed method. 

\begin{table}[ht]
\caption{WHEN network architecture. Size refers to either kernel shape or number of units. \#Fmaps is the number of feature maps in the layer. Activation denotes the activation used for the layer}
\centering
\begin{tabular}{lccc}
\toprule
\textbf{Layer}	& \textbf{Size}	& \textbf{\#Fmaps} & \textbf{Activation} \\
\midrule
Convolution 2D		& 3x3			& 64 & Linear \\
Batch Normalisation		& -			& - & - \\
Activation & - & - & ReLU \\ 
Max Pooling & 1x5 & - & - \\
Convolution 2D		& 3x3			& 64 & Linear \\
Batch Normalisation		& -			& - & - \\
Activation & - & - & ReLU \\ 
Max Pooling & 1x4 & - & - \\
Convolution 2D		& 3x3			& 64 & Linear \\
Batch Normalisation		& -	& - & - \\
Activation & - & - & ReLU \\ 
Max Pooling & 1x2 & - & - \\
Reshape & - & - & - \\
Bidirectional GRU & 64 & - & tanh \\
Bidirectional GRU & 64 & - & tanh \\
Time Distributed Dense & 64 & - & ReLU \\
Time Distributed Dense & 1 & - & Sigmoid \\
Flatten & - & - & - \\ 
\bottomrule
\end{tabular}
\label{tab:when}
\end{table}

The log mel-band energy feature extracted from the audio is fed to our network, which produces the predicted strong labels for each recording. The input to the proposed network is a $T$x40 feature matrix. The convolutional layers in the beginning of the network are in charge of learning the local shift-invariant features of this input. The max-pooling operation is performed along the frequency axis after every convolutional layer to reduce the dimension for the feature matrix while preserving the number of frames $T$. The output of the convolutional part of the network is then fed to  bi-directional gated recurrent units (GRUs) to learn the temporal structure of audio events. Next we apply time distributed dense layers to reduce feature-length dimensionality. Note that the time resolution of $T$ frames is maintained in both the GRU and dense layers. A sigmoid activation is used in the last time-distributed dense layer to produce a binary prediction of whether there is an event present in each time frame. This prediction layer outputs a continuous range of [0,1] for each frame of the input features. A prediction value equal or greater than 0.5 is taken to indicate the presence of the audio event in question while a prediction less than 0.5 signifies the absence of it. The dimensions of each prediction are $T$x1.

\section{Evaluation}
\label{sec:Eval}

\subsection{Datasets}
In order to test our approach in a low-resource setting we contacted our experiments on the training dataset provided during the Neural Information Processing Scaled for Bioacoustics (NIPS4B) bird song competition of 2013 and on a subset of the DCASE 2018 dataset used for Task 4: Large-scale weakly labeled semi-supervised sound event detection in domestic environments.\footnote{\url{http://sabiod.univ-tln.fr/nips4b/challenge1.html}}\footnote{\url{https://goo.gl/8wKkGk}} The first dataset contains birdsong recordings and the second one everyday domestic sound events.

The NIPS4B 2013 training set contains 687 recordings of maximum length of 5 seconds each. The recordings have already been weakly labelled with a total of 87 possible classes. Such a dataset can be considered low-resource for a few reasons. First, the total amount of training time is less than one hour. Also, there are 87 possible labels that have very sparse activations, 7 to 20 positive recordings for each. In order to efficiently use the data provided by the NIPS4B 2013 training dataset we first consider all 87 unique labels as one general label `bird' and train an audio event detection network for this audio event. Another limitation of this dataset is the imbalance of positive and negative recordings: out of the whole dataset (687 recordings) only 100 of them are labelled as negative. This caused some of our original experiments to classify almost all bags as positive ones. An easy way to solve the positive-negative imbalance is to force the network to have the same amount of positive and negative recordings in each mini-batch. Since the amount of negative recordings is much smaller compared to positive ones, we randomly repeat the negative recordings in each mini-batch. We refer to this as Half and Half (HnH) training. This provides a balanced training set at a bag level, but not necessarily balanced at an instance level.

For the following experiments, we split the NIPS4B 2013 training dataset into a training set and testing set. Only weak labels for the training set during the NIPS4B 2013 bird song competition were released, hence we could only use these recordings. We acquired the strong labels of most of these recordings via manual annotations, in order to use them for evaluating our system and have uploaded our transcriptions online.\footnote{\url{https://figshare.com/articles/Transcriptions_of_NIPS4B_2013_Bird_Challenge_Training_Dataset/6798548}} 

In order to evaluate how our method behaves with other types of audio events, we use a subset of the DCASE 2018 dataset of Task 4. In order to create a training set with less than an hour of training data that we can use as a low-resource dataset, we randomly select 360 recordings of maximum length of 10 seconds each out of the 1578 recordings of the task training set. DCASE consists of 10 classes. We combine three of them (Dog, Cat, Speech) into one general category named `mammals' and use that as our positive class for training our detector. Half and Half training is used in order to balance the amount of negative and positive recordings, that are originally 160 and 200 recordings, respectively. The full testing set of Task 4 is used for evaluating our method.

\begin{figure}[ht]
\centering
\begin{subfigure}[t]{0.4\textwidth}
   \includegraphics[width=1\linewidth]{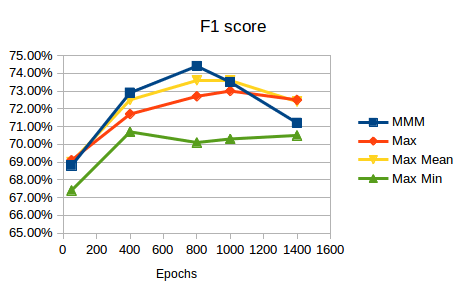}
   \caption{NIPS4B dataset}
   \label{fscore}
\end{subfigure}
\begin{subfigure}[t]{0.4\textwidth}
   \includegraphics[width=1\linewidth]{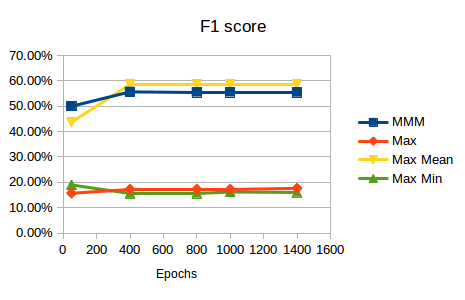}
   \caption{DCASE dataset}
   \label{fscore_dcase}
\end{subfigure}
\caption{Comparison of the progress of F1 score for our testing sets through epochs for different loss functions.}
\label{fig:metrics}
\end{figure}

\subsection{False Strong Labelling}

One of the most common ways of training networks from weak to strong labels is generating false strong labels for training by replicating the weak labels for every time frame of the audio and using them as strong labels. We train our network using false strong labels for 3000 epochs using binary cross-entropy as the loss function and Adam optimizer \cite{Kingma:14}. Figure \ref{623fsl} shows the resulting transcription predicted by the network. The main issue one can notice is that using false strong labels has a tendency of pushing all the results closer to one when dealing with a positive recording. Some structure is apparent in the transcription, hence the network is indeed able to differentiate between positive and negative instances to some degree, however all results for positive recordings are above the usual 0.5 threshold. We attribute this primarily to the nature of the false strong labels: for a positive recording all time frames are labelled as positive. Furthermore, since perfect accuracy for this setting does not correspond to the actual task, it becomes extremely unclear when training should be stopped.

\subsection{MIL using different loss functions}
We train two networks using the loss functions described in Section \ref{sec:bigidea}, namely max and MMM. Additionally, we trained two more networks using only the max and mean terms of MMM and the max and min terms of MMM to compare their performance and the impact each term has in the predictions of our audio event detector. 


Figures \ref{fscore} and \ref{fscore_dcase} present the progress of the F1 score, the harmonic average of the precision and recall of the predictions, on the NIPS4B and DCASE testing sets respectively, during training. One can notice that methods that use the mean term in the loss prediction tend to reach higher scores. For the NIPS4B dataset, we notice that after training for a certain amount of epochs the results for most methods are decreasing: this is due to the common issue of the MIL setting which is defining when one should stop training.


Another interesting aspect one can study is the individual results of the conventional max loss to our proposed MMM loss. Figure \ref{fig:max_mmm_623} depicts a positive recording from our NIPS4B testing set and the transcriptions predicted by each method. It is evident from these examples that once the max loss reaches the perfect accuracy for a bag, it ignores the harder-to-predict positive events. In this example, the network correctly predicts the three more prominent events and then ignores all other events between them. However, the network trained with a MMM loss is starting to pick out some of the harder to detect events, due to the gradient provided by using mean.

\begin{figure}
\centering
\begin{subfigure}[t]{0.33\textwidth}
   \includegraphics[width=1\linewidth]{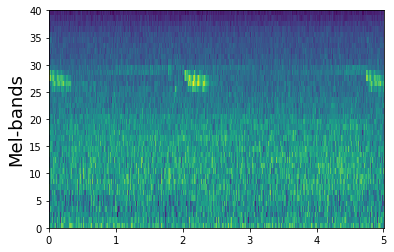}
\end{subfigure}\\
\begin{subfigure}[t]{0.33\textwidth}
	\includegraphics[width=1\linewidth]{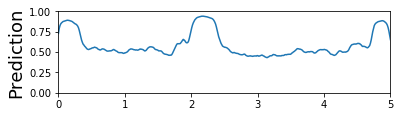}
	\caption{MIL using FSL}
	\label{623fsl}
\end{subfigure}
\begin{subfigure}[t]{0.33\textwidth}
   \includegraphics[width=1\linewidth]{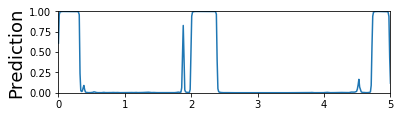}
   \caption{MIL using max}
   \label{623max}
\end{subfigure}\\
\begin{subfigure}[t]{0.33\textwidth}
   \includegraphics[width=1\linewidth]{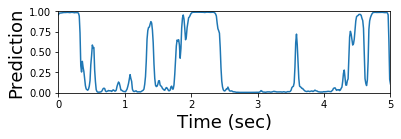}
   \caption{MIL using MMM}
   \label{623mmm}
\end{subfigure}
\caption{Predicted transcription of a recording from the testing set on the NIPS4B dataset. \ref{623fsl} depicts the results of our network trained in a false strong labelling setting. \ref{623max} depicts the results of our network trained with max loss. \ref{623mmm} depicts the results of our network trained with MMM loss.}
\label{fig:max_mmm_623}
\end{figure}
%

\vspace{-0.3cm}
\section{Conclusions}
\label{sec:discussion}
In this paper, we propose a method to perform audio event detection in a MIL setting that introduces a new loss function that takes into account predictions for all instances. Our method is tested on two low-resource datasets with only weakly labelled training data to perform bird and mammal vocalisation detection, respectively. We compare the different components of our loss function and define the influence each of them has to the training of the network. Training a network with our proposed loss outperforms the previously used losses in the MIL setting. Furthermore, our method is not tailored to any specific type of audio events, hence there is reason to believe it can be used for any kind of audio event detection tasks.
\bibliographystyle{IEEEtran}
\bibliography{stage2_bibliography}
%
%
%
%
%
%
%
%
%

\end{sloppy}
\end{document}